\documentclass[aps,prb,superscriptaddress,showpacs,twocolumn,longbibliography]{revtex4-2}

\usepackage{graphicx,amsmath,amssymb}
\usepackage{silence}
\usepackage[usenames]{color}
\usepackage{mathrsfs}
\usepackage{indentfirst}
\usepackage{float}
\usepackage{braket}
\usepackage[colorlinks=true, citecolor=blue, urlcolor=blue, linkcolor=blue ]{hyperref}
\usepackage{epstopdf}
\hypersetup{breaklinks=true}
\usepackage{booktabs}
\usepackage{lineno}

\begin{document}

	\title{Tracking doublon-holon dynamics in high-harmonic generation from Mott insulators}

	\author{Tao-Yuan Du }\email{duty710@163.com}  \affiliation{School of Mathematics and Physics, China University of Geosciences, Wuhan 430074, China}

	\author{Hui-Ru Li}   \affiliation{School of Mathematics and Physics, China University of Geosciences, Wuhan 430074, China}
	
  \author{Bo Li}   \affiliation{School of Mathematics and Physics, China University of Geosciences, Wuhan 430074, China}

    \author{Ruifeng Lu} \email{rflu@njust.edu.cn} \affiliation{ Institute of Ultrafast Optical Physics, Department of Applied Physics, Nanjing University of Science and Technology, Nanjing 210094, China}

    
	\begin{abstract}{
      High-harmonic generation (HHG) in strongly correlated Mott insulators is investigated using exact diagonalization and time-dependent density-matrix propagation of a laser-driven one-dimensional Hubbard chain. By projecting onto equilibrium Hubbard bands, we use the doublon population and its dynamics as a diagnostic to analyze intraband (spin-wave–like) and interband (doublon–holon creation) excitation channels. A filling-dependent crossover emerges: Bloch-like intraband response at dilute filling, mixed dynamics at intermediate filling, and interband-dominated HHG with plateau and cutoff near half filling. In the considered parameter range, increasing interaction strength $U$ strongly suppresses interband contributions through the enlarged Mott gap and correlation-induced localization. Intra- and interband current decomposition reveals opposing flows below the Mott gap ($\Delta_{\mathrm{Mott}}$) and selective dephasing suppression of interband coherence, enhancing net doublon accumulation. Time–frequency analysis uncovers the filling-dependent features of quantum trajectories, manifesting in distinct below-$\Delta_{\mathrm{Mott}}$ emission. This doublon-based analysis provides a transparent link between equilibrium spin–charge separation and nonequilibrium strong-field response, and clarifies how dephasing modifies interband coherence and doublon accumulation.}
		
	\end{abstract}.

	\maketitle

	\section{Introduction}

High-harmonic generation (HHG) in solids has emerged as a powerful probe of ultrafast quantum dynamics on sub-femtosecond timescales~\cite{Ferray1988,Corkum,goulielmakis2022high,heide2024ultrafast,silva,uzan2022observation,Ghimire2019,lattice}. In weakly correlated semiconductors and insulators, the HHG response admits a transparent single-particle Bloch interpretation: intraband emission stems from carrier acceleration within dispersive bands, while interband emission arises from excitation across a band gap followed by recombination~\cite{Vampa2014}. This dichotomy enables direct mapping of spectral features to band structure~\cite{uzan2022observation}. However, strong electron-electron correlations fundamentally invalidate the independent-quasiparticle picture~\cite{Yuta2,Shohei,Yuta2025,shaocan,Lange,Pizzi2023}, rendering such interpretations ambiguous in Mott insulators and related materials \cite{Udono,Masur,kskc-qlb3,m4x3-fb5d,Yuta2018,Markus,Yuta2021,Yuta2025,Yuta_RMP,arxiv2025}.

In strongly correlated systems, the relevant excitations are collective many-body states, naturally captured by the Hubbard model. Strong on-site repulsion $U$ splits the spectrum into lower and upper Hubbard bands (LHB/UHB) separated by a correlation-induced Mott gap $\Delta_{\mathrm{Mott}}$. Within this framework, intraband dynamics correspond to spin-wave--like motion within the LHB that conserves doublon number~\cite{Oka2012,Queisser2025,Murakami2022}, whereas interband dynamics involve doublon--holon pair creation across $\Delta_{\mathrm{Mott}}$, transferring spectral weight from spin- to charge-dominated sectors \cite{Yuta2,Yuta2025,Chang2024,Uchida}. The doublon density—the expectation value of on-site double occupancy—thus emerges as a physically transparent microscopic indicator: negligible for pure intraband motion, it clarifies interband charge excitation. Its filling dependence delineates distinct regimes: Bloch-like intraband response at dilute filling, mixed intra/interband dynamics at intermediate filling, and interband-dominated HHG with pronounced plateau and cutoff near half filling.

Previous studies of HHG in Mott systems have established the importance of doublon–holon excitations, particularly near half filling \cite{Silva2018,Yuta2025}. In contrast, the present work focuses on how the HHG mechanism evolves with filling level of electron, from dilute filling to the half-filled Mott-insulating regime. Using exact diagonalization and time-dependent density-matrix propagation \cite{huankai,Machao,temperature,Mrudul2020}, we analyze the time-dependent doublon population as a diagnostic quantity and relate it to the transition between intraband spin-wave–like motion and interband doublon–holon excitation. By classifying eigenstates according to doublon number, we analyze the relation between the equilibrium Hubbard-band structure and the nonequilibrium HHG response. This perspective enables us to identify a filling-dependent crossover, clarify the suppression of interband HHG with increasing on-site Coulomb repulsion in the considered range, and illustrate representative dephasing effects on interband coherence and doublon accumulation. While previous studies have established the role of doublon–holon excitations and current decomposition in HHG from Mott systems \cite{Yuta2018,Markus,Yuta2021,Yuta2025}, many of them focus primarily on the half-filled regime. Here we instead emphasize the filling dependence of the driven dynamics and use the time-dependent doublon population as a diagnostic quantity to track the crossover from single-particle intraband-dominated to doublon interband-dominated HHG.

This paper is organized as follows. In Sec.~\ref{sec2}, we present the theoretical model and computational methods based on the time-dependent density-matrix formalism applied to the one-dimensional Hubbard chain driven by a strong laser field. In Sec.~\ref{sec3}, we discuss the main results, including spectral features analyzed in terms of doublon dynamics, filling- and interaction-dependent dynamics, intra- and interband current decomposition, and the role of dephasing—including many-body effects—in doublon generation and harmonic emission. We conclude in Sec.~\ref{sec4}. Atomic units are used throughout unless specified otherwise.

\section{Theoretical model and methods} \label{sec2}

We consider the one-dimensional Hubbard model \cite{Essler2005,Mahan2000} as a paradigmatic framework for correlated electron systems, subject to periodic boundary conditions \cite{Hansen2022,Silva2018,Hansen2025}. To ensure spin neutrality, the numbers of spin-up and spin-down electrons are kept equal while varying the total particle number across different band fillings. The time-dependent Hamiltonian in the presence of an external laser field reads \cite{huankai,Silva2018}
\begin{align} \label{Eq1}
	\hat H(t) = & - t_0 \sum_{\sigma, j=1}^{L} 
	\left( e^{-i\phi(t)} \hat c_{j,\sigma}^\dagger \hat c_{j+1,\sigma} + \mathrm{H.c.} \right)  \nonumber \\
	&+ U\sum_{j=1}^{L} \hat n_{j\uparrow} \hat n_{j\downarrow},
\end{align}
where $\hat c^\dagger_{j\sigma}$ ($\hat c_{j\sigma}$) creates (annihilates) an electron with spin $\sigma=(\uparrow,\downarrow)$ on site $j$, and $\hat n_{j\sigma}=\hat c^\dagger_{j\sigma}\hat c_{j\sigma}$ is the local number operator. The hopping amplitude is set to $t_0=0.52~\mathrm{eV}$, and $U$ denotes the on-site Coulomb repulsion. The parameters used in the simulations are chosen to be broadly consistent with for quasi-one-dimensional cuprate chains such as Sr$_2$CuO$_3$ \cite{silva,Oka2012}. $U$ is varied to illustrate the trends and the general validity of our conclusions.

The electronic filling factor is defined as $n = \frac{N_\uparrow + N_\downarrow}{2L} \times 100\%$, where $L$ is the number of lattice sites. In the present half-filling calculations, we consider a chain with $L = 10$ sites under periodic boundary conditions. Here, $N_{\uparrow}$ and $N_{\downarrow}$ denote the numbers of spin-up and spin-down electrons, respectively. For a given filling, the Hilbert-space dimension is $\binom{L}{N_{\uparrow}} \times \binom{L}{N_{\downarrow}}$, where $\binom{L}{N}$ denotes the number of ways to choose $N$ sites out of $L$. Since the dynamics considered here are governed primarily by local doublon--holon correlations, the results are expected to remain qualitatively robust for larger system sizes. A half-filled system ($n$ = 50\%) corresponds to one electron per site on average, while $n < 50\%$ represents hole-doped configurations. Note that for the filling cases of 5\% and 25\%, different chains are considered: a chain with $L = 20$ containing one electron pair, and a chain with $L = 8$ containing two electron pairs. The time-dependent Peierls phase is $\phi(t)=a_0A(t)$, where $a_0=7.56$~a.u. is the lattice constant of the adopted system, and the electric field, $F(t)=-dA(t)/dt$, is related to the vector potential $A(t)$. The driving field is taken as a linearly polarized pulse with a $\sin^2$ envelope, a total duration of ten optical cycles, and a carrier wavelength of 9.11 $\mu$m. This driving laser wavelength lies in the mid-infrared regime commonly employed in strong-field HHG experiments in solids \cite{silva,Oka2012}.

The field-free Hamiltonian $\hat H_0 \equiv \hat H(t)|_{\phi(t= 0) }$ is diagonalized in the configuration (Fock) basis, $\hat H_0|\alpha\rangle=E_\alpha|\alpha\rangle$. The unitary transformation \small{$V=[|\alpha\rangle]_{\alpha=1}^M \in \mathbb{C}^{M\times M}$} satisfies $V^\dagger V = VV^\dagger = \mathbb{I}$, allowing any operator $\hat O$ to be represented in the many-body eigenbasis as $\tilde O = V^\dagger \hat O V$. The doublon operator, which measures on-site double occupancy, is defined as $	\hat N_d=\sum_{i=1}^L \hat n_{i\uparrow} \hat n_{i\downarrow}$, and its state-resolved expectation value $d_\alpha=\langle \alpha| \hat N_d|\alpha\rangle$ quantifies the doublon content of each eigenstate. The electric current operator in the lattice basis is given by
\begin{equation} \label{Eq2}
	\hat J(t) = -i a_0 t_0 \sum_{\sigma}\sum_{j=1}^L 
	\left( e^{-i\phi(t)} \hat c_{j,\sigma}^\dagger \hat c_{j+1,\sigma} - \mathrm{H.c.} \right).
\end{equation}

Quantum decoherence in the many-body wave-packet dynamics is incorporated via a pure-dephasing term in the von Neumann equation within a Lindblad-type master-equation framework \cite{Machao}. In the eigenbasis, the density-matrix evolution reads
\begin{equation}  \label{Eq3}
	\dot{\rho}_{mn}(t) = -i\big[\tilde H(t),\rho(t)\big]_{mn} 
	- \frac{1-\delta_{mn}}{T_2}\rho_{mn}(t),
\end{equation}
where $\tilde H(t)=V^\dagger\hat H(t)V$. This form explicitly shows that off-diagonal coherences ($m\neq n$) decay exponentially with dephasing time $T_2$, while populations ($m=n$) remain conserved. Here $T_2$ is introduced phenomenologically to illustrate the role of coherence loss in the driven dynamics, rather than to model a specific material quantitatively. Real-time propagation is performed using a fourth-order Runge–Kutta algorithm. The time step used in the propagation is $\Delta t$ = 0.1 a.u. Convergence with respect to $\Delta t$ was verified by comparing the current and doublon dynamics for smaller time steps, for which no visible differences were found in the reported spectra.

 \begin{figure}
		\includegraphics[width= 8.9 cm,  height = 6 cm]{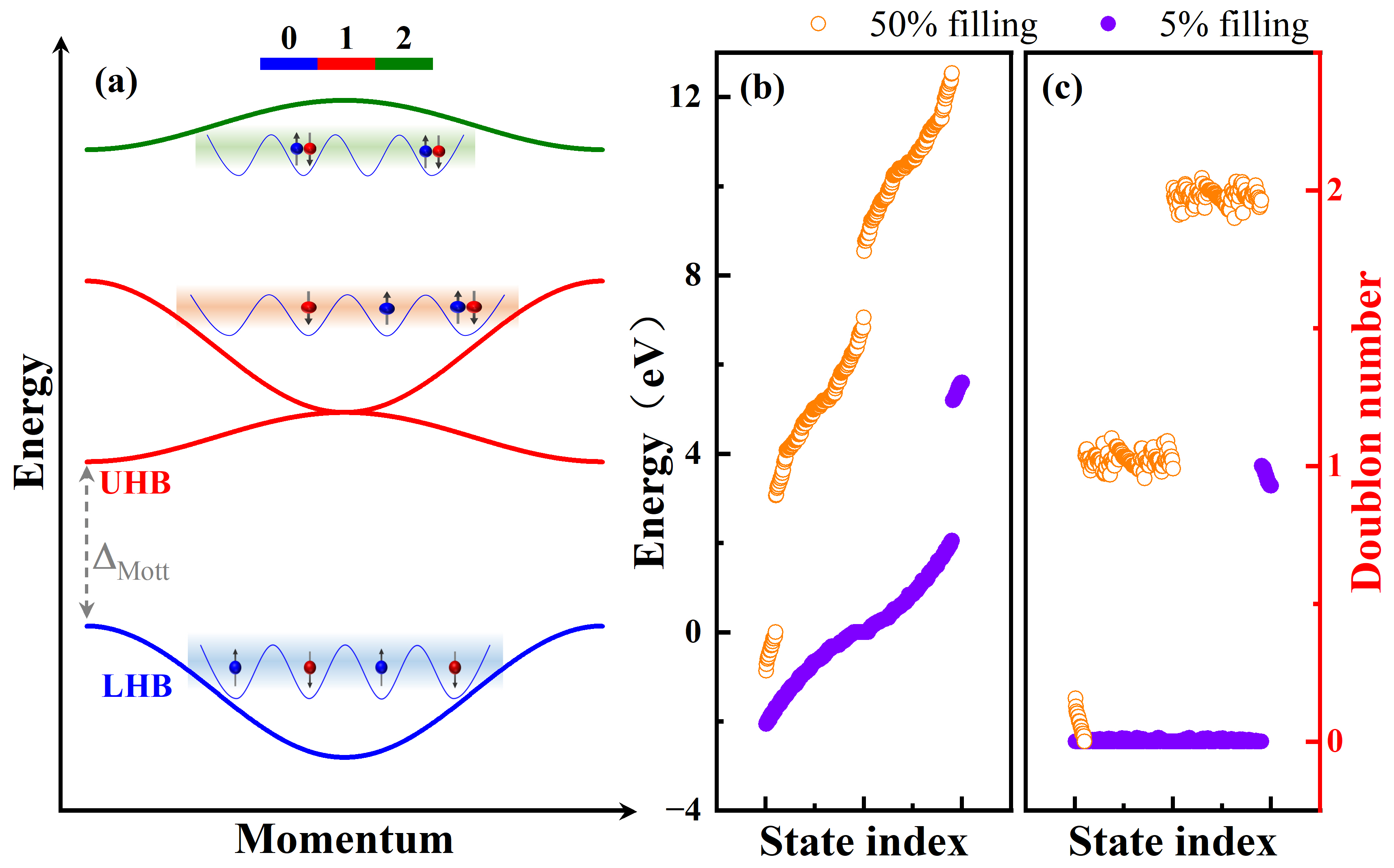}
		\caption{(a) Quasiparticle dispersions extracted from equilibrium spectral-function analysis, color-coded by the doublon expectation value. Insets show representative real-space occupations. The separation between the lower and upper Hubbard bands defines the Mott gap $\Delta_{\mathrm{Mott}}$. In the many-body ground state, the lower band is dominated by singly occupied (spin-wave–like) states, whereas doublon occupations mainly populate the upper band. Increasing the filling enhances the doublon weight, signaling the formation of multi-doublon configurations. (b) Filling-dependent eigenenergies for $U/t_0$ = 10 and (c) their corresponding doublon occupations of eigenstates.}
		\label{Fig1}
	\end{figure}

\begin{figure*}
		\includegraphics[width =  0.7 \textwidth]{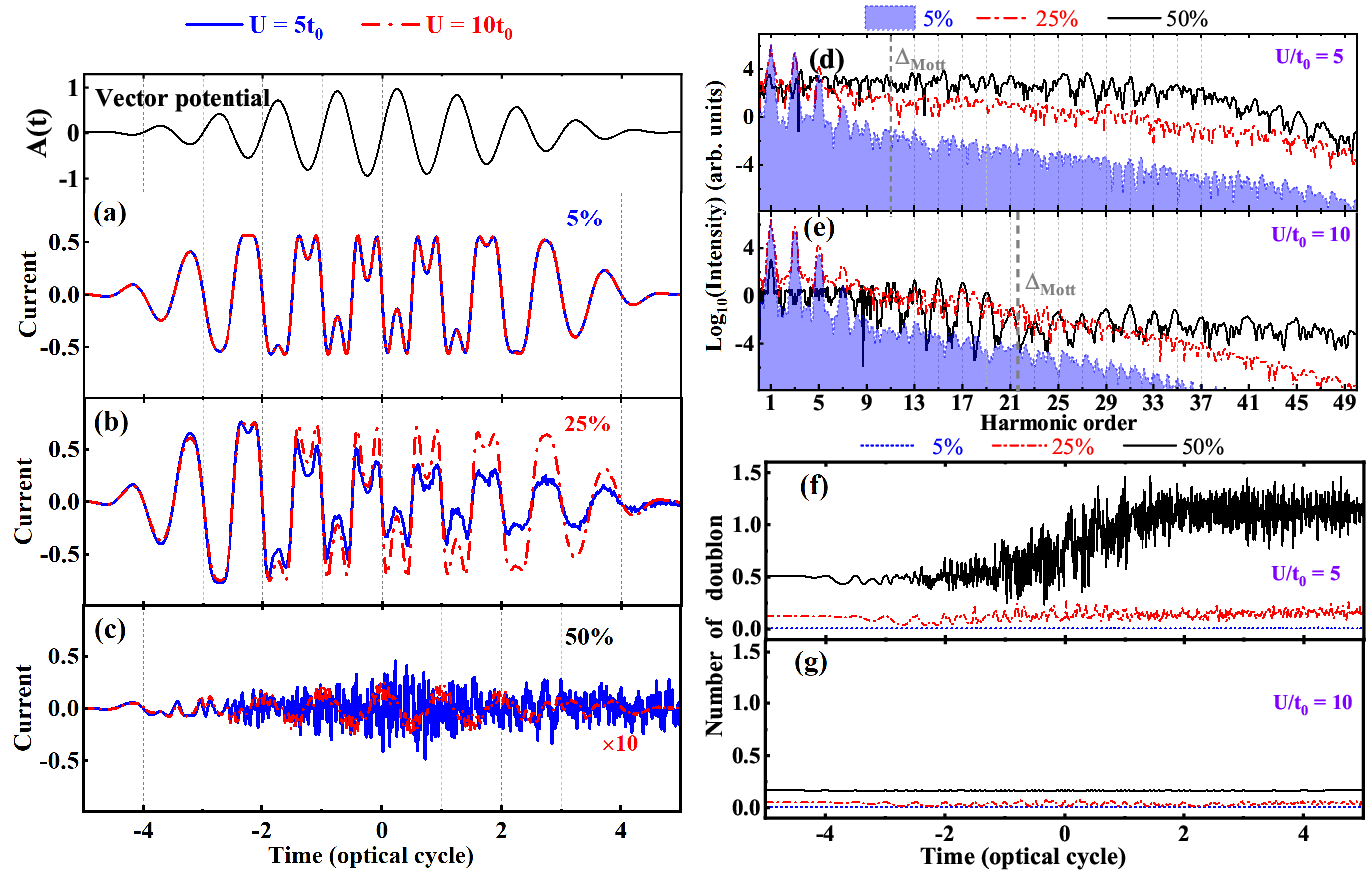}
		\caption{Time-dependent current of a one-dimensional Hubbard chain driven by a laser pulse with the peak field strength of 10~MV/cm, comparing on-site Coulomb repulsion values of $U=5t_0$ (blue solid) and $U=10t_0$ (red dashed) at three electronic fillings: (a) dilute (5\%), (b) intermediate (25\%), and (c) half filling (50\%). Panels (d)–(e) display the corresponding high-harmonic spectra, while panels (f)–(g) show the time evolution of the doublon population $D(t)$ under the same conditions.}
		\label{Fig2}
\end{figure*}

The initial state $\rho(0)$ is typically chosen as the equilibrium density matrix of $\hat H_0$ (e.g., its ground state). Observables are evaluated in the eigenbasis: with $\tilde J(t)=V^\dagger\hat J(t)V$ and $\tilde N_d=V^\dagger\hat N_dV$, we compute the expectation values of the current $J(t)=\operatorname{Tr}[\tilde J(t)\rho(t)]$ and the doublon population $D(t)=\operatorname{Tr}[\tilde N_d\rho(t)]$. To gain deeper insight into the intra- and interband dynamics of doublon–holon pairs, we classify the many-body eigenstates of Fig. \ref{Fig1}(b) into effective energy bands, labeled by index $l$, according to their doublon-weight distributions in Fig. \ref{Fig1}(c). Under the density-matrix framework \cite{Machao}, we introduce band projection operators $P_l = \sum_{\alpha \in B_l} |\alpha\rangle\langle\alpha|$, where $B_l$ denotes the index set of eigenstates belonging to band $l$. The total intraband and interband projectors are defined as $P_{\text{intra}} = \sum_l P_l$ and $P_{\text{inter}} = \mathbb{I} - P_{\text{intra}}$, respectively, with $\mathbb{I}$ being the identity operator. The density matrix is then partitioned into $\rho_{\text{intra}} = P_{\text{intra}}\rho P_{\text{intra}}$ and $\rho_{\text{inter}} = \rho - \rho_{\text{intra}}$. Accordingly, the total current $J(t) = \mathrm{Tr}[\hat{J}\rho]$ can be exactly decomposed into intraband and interband components, $J_{\text{intra}} = \mathrm{Tr}[\hat{J}\rho_{\text{intra}}]$ and $J_{\text{inter}} = \mathrm{Tr}[\hat{J}\rho_{\text{inter}}]$, respectively. The HHG spectrum is obtained from the Fourier transform of currents.

	\section{Results and discussion} \label{sec3}

\begin{figure*}
		\includegraphics[width = 0.8 \textwidth ]{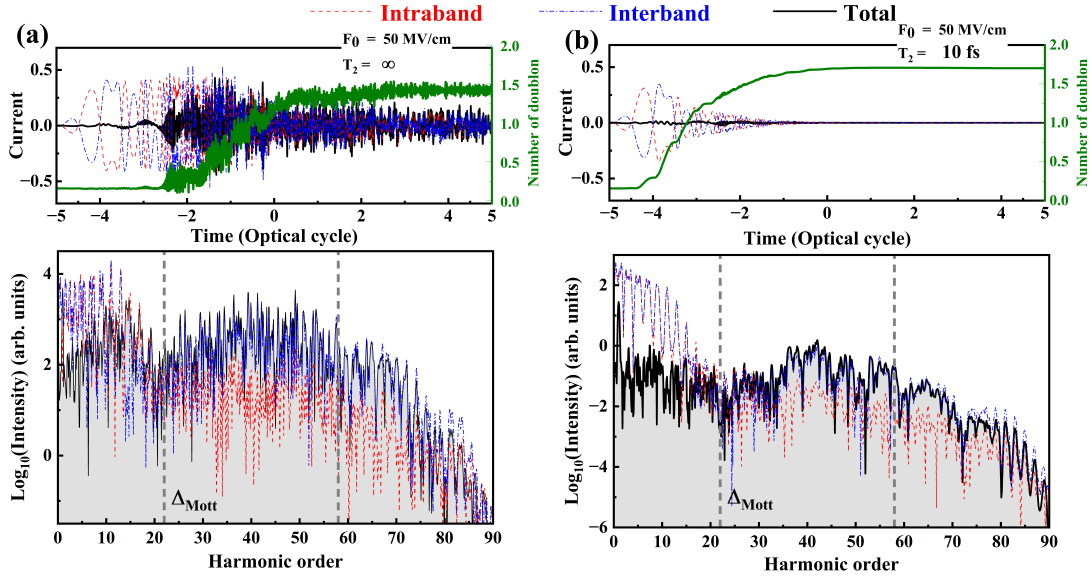}
		\caption{Separation of intra- and interband current contributions for the half-filled case with $U/t_0 = 10$ under strong-field driving. Panels (a) and (b) correspond to laser fields with a peak strength of 50~MV/cm, above the excitation threshold, shown without and with dephasing, respectively. The olive-green curves indicate the corresponding time-dependent doublon populations $D(t)$ for the two cases. } 
		\label{Fig3}
\end{figure*}

	In equilibrium, the spectral function analyzed together with the doublon expectation value separates spin and charge degrees of freedom, as illustrated in Fig.~\ref{Fig1}. Based on the intrinsic correspondence between the eigenstates and their associated doublon numbers illustrated in Fig. ~\ref{Fig1}, we establish a clear characterization of the effective bands. The lower Hubbard band is dominated by spin-wave–like singly occupied states, while the upper Hubbard band carries finite doublon weight and reflects charge-density fluctuations. In Figs. \ref{Fig1}(a) and \ref{Fig1}(b), the separation between these bands defines the Mott gap $\Delta_{\text{Mott}}$, which sets the energetic threshold for doublon–holon creation. Under strong laser driving, this spin–charge framework dictates the nonequilibrium response: intraband motion within the lower Hubbard band corresponds to quantized spin-wave excitations that conserve doublon number and predominantly generate low-order harmonics, whereas interband transitions across the Mott gap create doublon–holon pairs, transfer spectral weight from spin-wave–dominated to charge-density–dominated states, and produce the spectral plateau and cutoff characteristic of HHG. In this sense, the doublon numbers of eigenstates in Fig. \ref{Fig1}(c) characterize the doublon dynamics, which provides a useful microscopic perspective on the nonlinear current response and helps connect equilibrium correlations to the strong-field dynamics.

Figure~\ref{Fig1}(c) shows the doublon number associated with each eigenstate. The distinct doublon distributions under the two filling conditions imply that the corresponding light-driven doublon dynamics lead to different harmonic radiation. In the following, we analyze the filling-dependent doublon dynamics and their relation to the nonlinear current and HHG spectra. Figures~\ref{Fig2}(a-c) then show the filling-dependent current dynamics. The three fillings considered here (5\%, 25\%, and 50\%) are chosen as representative cases of dilute, intermediate, and half-filled regimes, respectively, to illustrate the crossover of the HHG mechanism with filling. At dilute filling (5\%, Fig.~\ref{Fig1}(a)), the current follows the pulse envelope and exhibits nearly sinusoidal oscillations, reflecting intraband Bloch motion within the lower Hubbard band and spin-wave precession. The responses for $U=5t_0$ and $U=10t_0$ are nearly indistinguishable, consistent with the negligible doublon probability predicted from equilibrium. At intermediate filling (25\%, Fig.~\ref{Fig2}(b)), the response exhibits quasi-periodic oscillations during the initial cycles, after which driving carriers across the Mott gap gives rise to higher-frequency components. This onset of interband excitation is more pronounced for smaller $U/t_0$, reflecting the lower doublon–holon threshold and enhanced interband polarization. Near the peak of the laser envelope, doublons are efficiently generated; however, because the upper Hubbard band is narrow energy band dispersion shown in Fig. \ref{Fig1}(a) and its group velocity small, the current amplitude decreases once doublons accumulate. At half filling (50\%, Fig.~\ref{Fig2}(c)), interband processes dominate: for $U=5t_0$, broadband high-frequency currents appear, whereas for $U=10t_0$, the response is strongly suppressed due to correlation-induced localization and a higher excitation threshold.

The harmonic spectra corroborate the filling-dependent behaviors in above-mentioned currents. For dilute filling (5\%) at $U=5t_0$, the spectrum decays rapidly with increasing harmonic order and exhibits enhanced odd low-order peaks, as shown by the blue-shaded spectrum in Fig.~\ref{Fig2}(d). In this regime, the doublon number $D(t)$ remains nearly constant and close to zero [blue dashed curve in Fig.~\ref{Fig2}(f)], confirming that the response is dominated by intraband dynamics. At intermediate filling (25\%), the harmonic yield, represented by the red dash-dotted curve in Fig.~\ref{Fig2}(d), is substantially enhanced relative to the dilute case, marking the onset of doublon–holon generation. At half filling (50\%), strong high-harmonic emission emerges at the HHG plateau, as indicated by the black solid curve in Fig.~\ref{Fig2}(d), while the time-dependent doublon number $D(t)$ increases rapidly before saturating [black solid curve in Fig.~\ref{Fig2}(f)], consistent with interband excitation saturation and persistent doublon–holon recombination that sustains an extended HHG plateau. Increasing the on-site interaction to $U=10t_0$ enlarges the Mott gap and suppresses light-induced doublon excitation to higher-lying bands, thereby leading to a systematic reduction of interband contributions across all fillings. As shown in Fig.~\ref{Fig2}(e), the spectra are uniformly weakened and the filling-induced plateau enhancement observed at $U=5t_0$ is strongly reduced. Consistently, the doublon dynamics in Fig.~\ref{Fig2}(g) show negligible growth for dilute filling, strongly inhibited growth for intermediate filling, and a nearly flat profile at half filling. In the present simulations, stronger on-site Coulomb repulsion suppresses doublon–holon creation and weakens interband HHG channels across fillings. This behavior is qualitatively consistent with a correlation-induced doublon blockade, whereby stronger interactions reduce charge mobility and limit spin–charge energy transfer. In addition, as shown in Figs.~\ref{Fig2}(d) and \ref{Fig2}(e), the harmonic peaks become progressively clearer with increasing filling, indicating a change in the underlying emission dynamics \cite{vampa_semi}, which will be discussed further below.

To classify the contributions of intra- and interband dynamics of doublon–holon pairs, Fig.~\ref{Fig3} presents the decomposition of intra- and interband currents for the half-filled case with $U/t_0 = 10$. As indicated by the finite doublon number within the lowest Hubbard band (orange markers in Fig.~\ref{Fig1}(c)), the eigenstates of this band already possess a non-negligible doublon number. Consequently, the intraband dynamics deviate from the strict picture of purely doublon-conserving motion, reflecting the eigenstate-resolved doublon number within the lower Hubbard band. Thus, when the quantum many-body state within the lower Hubbard band carries a non-negligible doublon number, the resulting dynamics involve the motion of doublons between lattice sites, which in turn leads to characteristics typical of interband processes. This explains why, in the HHG spectrum of Fig.~\ref{Fig3}, the harmonic intensities below $\Delta_{\mathrm{Mott}}$ are comparable in magnitude for the intra- and interband contributions, while the corresponding currents flow in opposite directions.

When dephasing is introduced, the coherence time $T_2$ determines how rapidly the off-diagonal elements of the density matrix $\rho_{\mathrm{inter}}(t)$ decay. Finite dephasing selectively suppresses interband coherences, thereby reducing $J_{\mathrm{inter}}(t)$ and weakening the high-harmonic plateau. Interestingly, the total doublon population at the end of the laser pulse increases, consistent with observations in uncorrelated systems \cite{Vampa2014,Wanggan,Du2021,Zhang2025}. This occurs because dephasing inhibits coherent Rabi oscillations between doublon–holon states and prevents population from returning to the ground sector. As a result, doublons accumulate progressively over the entire laser pulse, as shown in the upper panels of Figs. \ref{Fig3}(a,b). In parallel, dephasing slightly reduces the amplitude of $J_{\mathrm{intra}}(t)$ and alleviates the destructive interference between $J_{\mathrm{intra}}$ and $J_{\mathrm{inter}}$ in the sub-Mott-gap region. Consequently, the harmonic yield below $\Delta_{\mathrm{Mott}}$ becomes smoother and exhibits a higher signal-to-noise ratio, characterized by more distinct low-order harmonic peaks. This behavior demonstrates that dephasing plays a dual role in correlated electron dynamics: it suppresses interband coherence while enhancing net doublon generation by disrupting coherent population exchange, and it simultaneously reshapes the interference between intra- and interband currents.

Microscopically, dephasing acts as a dynamical filter that selectively damps quantum coherences across the Mott gap while partially preserving intraband carrier motion within the lower Hubbard band. As a result, the system exhibits a tendency from coherence-dominated dynamics toward more incoherent behavior, in which doublon–holon pairs undergo reversible oscillations, to an incoherent regime characterized by population relaxation and quasi-classical transport, where doublon accumulation and energy dissipation become dominant. These effects are experimentally accessible through high-harmonic generation measurements. By tuning environmental parameters such as temperature, disorder, or electron–phonon coupling, which effectively govern the coherence time $T_2$, one can systematically modulate the balance between the high-harmonic plateau and the sub-gap response. Specifically, shorter coherence times are expected to suppress high-order harmonics, enhance low-order peaks, and increase the residual doublon population, reflecting the crossover from coherent to incoherent dynamics \cite{Hogger,Murakami2022,orthodoxou2021}. Complementary time-resolved spectroscopies, such as pump–probe photoemission or transient optical absorption, could directly probe this doublon accumulation and the evolution of interband coherence. Overall, dephasing not only governs the lifetime of interband quantum coherence but also regulates doublon–holon generation and the interference landscape of strong-field dynamics, underscoring its central role in controlling the balance between coherent and incoherent contributions in driven Mott systems.

\begin{figure}
	\includegraphics[width = 9 cm, height = 3.5 cm ]{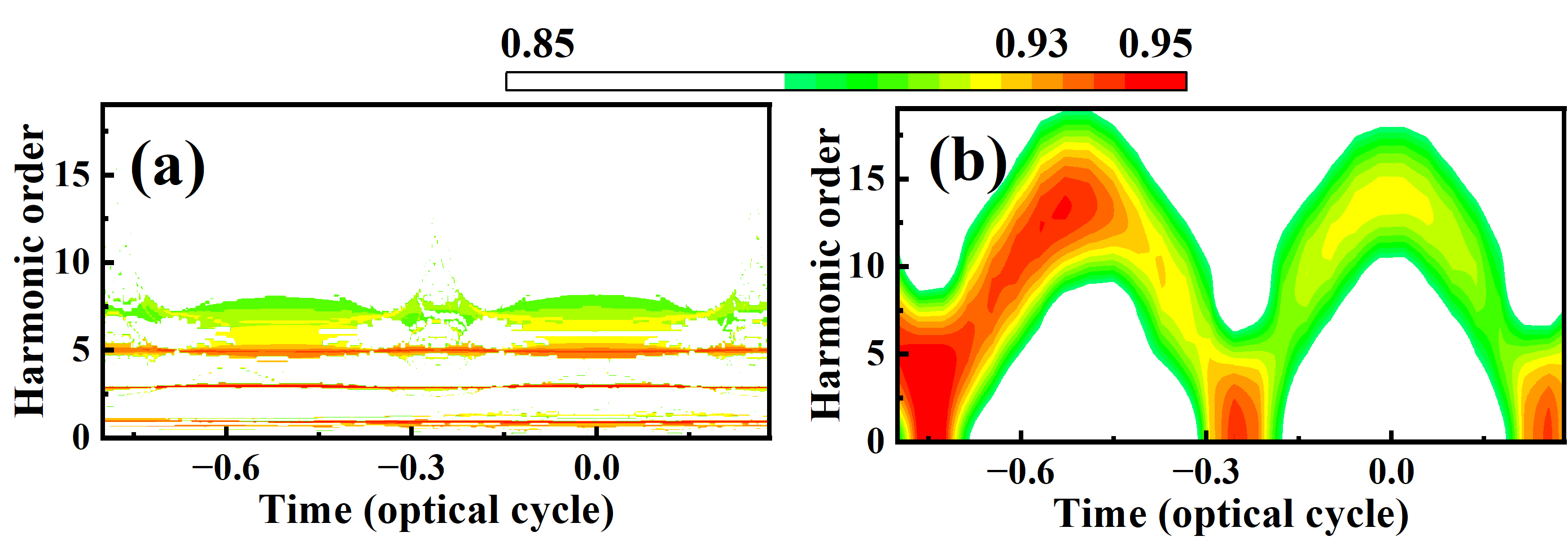}
	\caption{Quantum-trajectory features of the below-$\Delta_{\mathrm{Mott}}$ spectral region with varying electron filling. Panels (a) and (b) correspond to the dilute (5\%) and half-filling (50\%) cases, respectively. The HHG spectra are obtained with parameters $U/t_0 = 10$ and a peak electric-field strength of 50 MV/cm. The color scale of the intensity is independently normalized for each spectral region.}
	\label{Fig4}
\end{figure}

Finally, we examine the influence of electron filling on the quantum trajectories of the below-$\Delta_{\mathrm{Mott}}$ high-harmonic generation. To ensure a rigorous analysis, the pure-dephasing term in Eq.~(\ref{Eq3}) is excluded from the simulation, allowing us to isolate and examine the intrinsic doublon number of the eigenstate arising solely from electron filling. Figure~\ref{Fig4} presents the time–frequency analysis of HHG for the dilute (5\%) and half-filling (50\%) cases. In Fig.~\ref{Fig4}(a), the temporal emission characteristics in the below-$\Delta_{\mathrm{Mott}}$ region exhibit time–frequency features associated with below-gap harmonic emission contributed by Bloch electrons, which do not display distinct quantum trajectories~\cite{Wumengxi,vampa_semi}. A clear distinction emerges between the two filling cases: the dilute case (5\%) exhibits Bloch-like intraband oscillations~\cite{Wumengxi}, whereas the half-filled case (50\%) gives rise to well-defined quantum emission trajectories in Fig.~\ref{Fig4}(b). This difference indicates that the emergence of quantum trajectories originates from doublon dynamics, specifically the recombination of doublons and holons, which is enabled by the presence of doublon occupation within the eigenstates of the lower Hubbard band at half filling (50\%), in contrast to its absence in the dilute case (5\%), as illustrated in Fig.~\ref{Fig1}(c). In summary, to exhibit clear quantum trajectories in harmonic emission within the below-$\Delta_{\mathrm{Mott}}$ region, a non-zero doublon number of many-body eigenstates appears to favor the emergence of such trajectory features. This also implies that the many-body eigenstates of the lower Hubbard band, possessing a non-zero doublon number and oscillating within the band, are fundamentally different from the Bloch oscillation of a single-electron state. In addition, the quantum trajectory analysis in Fig.~\ref{Fig4} shows weakened longer trajectory features at high filling, implying many-body-blockade-like suppression of long-range doublon motion.

Overall, the present results reveal three robust trends in the driven Hubbard system. First, the HHG response evolves from single-electron intraband-dominated emission at dilute filling to doublon–holon recombination interband-dominated plateau formation near half filling. Second, the growth of the doublon population correlates directly with the emergence of the HHG plateau, providing a microscopic indicator of doublon generation. Third, increasing the on-site Coulomb interaction suppresses interband HHG in the considered range by enlarging the Mott gap and enhancing correlation-induced localization.

    \section{Conclusion} \label{sec4}
We have presented a microscopic analysis of high-harmonic generation in correlated electron systems based on doublon–holon dynamics. Using time-dependent density-matrix simulations of the driven one-dimensional Hubbard model, we show that the doublon yield serves as a useful microscopic indicator of the balance between intraband spin-wave motion and interband doublon–holon excitation. Its evolution captures a filling-dependent crossover—from Bloch-like intraband response at low filling (5\%), through mixed dynamics at intermediate filling (25\%), to interband-dominated behavior near half filling (50\%)—and a strong suppression of interband HHG with increasing on-site Coulomb interaction, reflecting the enlarged Mott gap and correlation-induced localization. Rigorous decomposition of the total current into intra- and interband components reveals that finite dephasing selectively suppresses interband coherence, enhances net doublon accumulation, and mitigates destructive interference below $\Delta_{\mathrm{Mott}}$. Time-frequency analysis further suggests that many-body-blockade-like suppression of long-range doublon motion contributes to the weakening of longer trajectory features at high filling. These results position doublon–holon dynamics as a physically transparent and experimentally accessible probe, bridging equilibrium spectral correlations with nonequilibrium strong-field transport and offering a viable route for understanding and potentially controlling nonlinear optical responses in Mott materials.

\section*{ACKNOWLEDGMENTS}
R.L. is supported by National Key R\&D Program of China (2022YFA1604301), National Natural Science Foundation of China (12425411), the Natural Science Foundation of Jiangsu Province (BK20253027),  and the Funding of NJUST (TSXK2022D003).

\section*{DATA AVAILABILITY}
The data that support the findings of this article are not publicly available. The data are available from the authors upon reasonable request.


%

\end{document}